\documentclass[conference]{IEEEtran}
\usepackage{hyperref} 
\pdfoutput=1 
\usepackage{epsfig}
\usepackage{graphicx}
\usepackage{algorithm}

\usepackage{ifpdf}

\usepackage{cite}
\usepackage{amsmath, amssymb}
\usepackage{algorithmic}

\usepackage{array}

\usepackage{mdwmath}

\begin{document}
%
\title{ Data Modeling with Large Random Matrices in a Cognitive Radio Network Testbed: Initial Experimental Demonstrations with 70 Nodes}


 \author{\IEEEauthorblockN{Changchun Zhang and  Robert C. Qiu
 }
\IEEEauthorblockA{Cognitive Radio Insitute,
Department of Electrical and Computer Engineering, \\Center for Manufacturing Research, Tennessee Technological University, \\Cookeville, Tennessee 38505, USA\\
Email: czhang42@students.tntech.edu, rqiu@tntech.edu}

 }

\maketitle

\begin{abstract}
This short paper reports some initial experimental demonstrations of the theoretical framework: the massive amount of data in the large-scale cognitive radio network can be naturally modeled as (large) random matrices. In particular, using experimental data we will demonstrate that the empirical spectral distribution of the large sample covariance matrix---a Hermitian random matrix---agree with its theoretical distribution (Marchenko-Pastur law). On the other hand, the eigenvalues of the large data matrix ---a non-Hermitian random matrix---are experimentally found to follow the single ring law, a theoretical result that has been discovered relatively recently. To our best knowledge, our paper is the first such attempt, in the context of large-scale wireless network, to compare theoretical predictions with experimental findings.
\end{abstract}

\section{Introduction}

Statistical science is an empirical science.  The object of statistical methods, according to R.A. Fisher (1922)~\cite{fisher1922mathematical}, is the reduction of data: ``It is the object of the statistical processes employed in the reduction of data to exclude this irrelevant information, and isolate the whole of the relevant information contained in the data.'' In the age of Big Data~\cite{Qiu_WicksBook2013,QiuAntonik2014Wiley}, the goal set by Fisher has never been so relevant today. In the work of~\cite{QiuBook2012Cognitive}, we make an \textit{explicit} connection of big data with large random matrices. This connection is based on the simple observation that massive amount of data can be naturally represented by (large) random matrices. When the dimensions of the random matrices are sufficiently large, some unique phenomena (such as concentration of spectral measure) will occur~\cite{Qiu_WicksBook2013,QiuAntonik2014Wiley}. 

In the same spirit of our previous work---representing large datasets in terms of random matrices, we report some empirical findings in this short paper. In this initial report, we summarize the most interesting results only when the theoretical models agree with experimental data. When the size of a random matrix is sufficiently large, the empirical distribution of the eigenvalues (viewed as functions of this random matrix) converges to some theoretical limits (such as Marchenko-Pastur law and the single ring law). In the context of large-scale wireless network, our empirical findings will validate these theoretical predictions. To our best knowledge, our work represents the first such attempt in the literature, although a lot of simulations are used in the past work~\cite{CouilletDebbah2011BookRMT}. 


The structure of this paper is as follows. We first describe the theoretical models. Then, empirical findings are compared with these theoretical models.

\section{Theoretical Statistical Models}
\subsection{Marchenko-Pastur Law}
\label{sect:MPlaw}
Let ${\bf X} = {\left\{ {{\xi _{ij}}} \right\}_{1 \leqslant i \leqslant N,1 \leqslant j \leqslant n}}$ be a random $N \times n$ matrix whose entries are i.i.d. $N$ is an integer such that $N \leqslant n$ and $N/n = c$ for some $c \in \left( {0,1} \right]$. The empirical spectrum density (ESD) of the corresponding sample covariance matrix ${\bf S} = \frac{1}{n}{{\bf X}^H}{\bf X}$ converges to the distribution of Marchenko-Pastur law~\cite{CouilletDebbah2011BookRMT,QiuBook2012Cognitive} with density function 
\begin{equation}
{f_{MP}}\left( x \right) = \left\{ {\begin{array}{*{20}{c}}
  {\frac{1}{{2\pi xc{\sigma ^2}}}\sqrt {(b - x)(x - a)} ,\quad \quad a \leqslant x \leqslant b} \\ 
  {0\quad {\kern 1pt} \quad \quad \quad \quad \quad \quad \quad \quad \quad \quad otherwsie} 
\end{array}} \right.
\label{eq:(MPlaw}
\end{equation}
where $a = {\sigma ^2}{(1 - \sqrt c )^2},\quad b = {\sigma ^2}{(1 + \sqrt c )^2}.$

\subsection{Kernel Density Estimation}
A nonparametric estimate~\cite{pan2011universality} of the empirical spectral density of the sample covariance matrix $\bf S$ can be used
\begin{equation}
{f_n}(x) = \frac{1}{{ph}}\sum\limits_{i = 1}^p {K(\frac{{x - {\lambda _i}}}{h})} 
\end{equation}
in which ${\lambda _i},\;i = 1,...,p,$ are the eigenvalues of $\bf S$, and $K(\cdot)$ is the kernel function for bandwidth parameter $h.$

\subsection{Power Law}
Sometimes, the empirical spectral density can be fitted by a power law which is given by
\begin{equation}
\begin{split}
\rho _{\beta }\left ( x \right ) = &\frac{1}{2\pi c\beta \Gamma \left ( \beta +1 \right )}\left (\frac{c\beta}{x} \right )^{\beta +2}. \\
&\int_{X_{-}}^{X_{+}}t^\beta \textup{exp}\left. ( -\frac{c\beta }{x} \right. t)\sqrt{\left ( t-X_{-} \right )\left ( X_{+} -t \right )}dt,
\end{split}
\end{equation}
where $\beta$ is the power law parameter that will be experimentally found via curve fitting. 
The power law is mainly applicable for the case when signal is present.
	
\subsection{The Single ``Ring'' Law}

Let, for each $n\ge 1,$ ${\bf A}_n$ be a random matrix which admits the decomposition ${{\mathbf{A}}_n} = {{\mathbf{U}}_n}{{\mathbf{T}}_n}{{\mathbf{V}}_n},$ with ${{\mathbf{T}}_n} = \operatorname{diag} \left( {{s_1},...,{s_n}} \right)$ where the $s_i$'s are positive numbers and where ${{\mathbf{U}}_n}$ and  ${{\mathbf{V}}_n}$ are two independent random unitary matrices which are Haar-distributed independently from the matrix ${{\mathbf{T}}_n}.$ Under certain mild conditions, the ESD ${\mu _{{{\mathbf{A}}_n}}}$ of ${\bf A}_n$ converges~\cite{guionnet2009single}, in probability, weakly a deterministic measure  whose support is $\left\{ {z \in \mathbb{C}:a \leqslant \left| z \right| \leqslant b} \right\},$ $a = {\left( {\int {{x^{ - 2}}\nu \left( {dx} \right)} } \right)^{ - 1/2}},b = {\left( {\int {{x^2}\nu \left( {dx} \right)} } \right)^{1/2}}.$ Some outliers to the single ring law~\cite{benaych2013outliers} can be observed. 

Consider the matrix product $\prod\limits_{i = 1}^\alpha  {{{\bf X}_{i}}},$ where ${{\bf X}_{i}}$ is the singular value equivalent~\cite{cakmak2012non} of the rectangular $N \times n$ non-Hermitian random matrix ${\tilde{\bf X}_i},$ whose entries are i.i.d.  Thus, the empirical eigenvalue
distribution of $\prod\limits_{i = 1}^\alpha  {{{\bf X}_{i}}}$ converge almost surely to the same limit given by 
\begin{equation}
{f_{\prod\limits_{i = 1}^\alpha  {{{\bf X}_{i}}} }}(z) = \left\{ {\begin{array}{*{20}{c}}
  {\frac{1}{{\pi c\alpha }}{{\left| z \right|}^{2/\alpha  - 2}}}&{{{(1 - c)}^{\alpha /2}} \leqslant \left| z \right| \leqslant 1} \\ 
  0&{elsewhere} 
\end{array}} \right.
\label{circular_law}
\end{equation}
as $N,n \to \infty $ with the ratio $c = {N \mathord{\left/
 {\vphantom {N n}} \right.
 \kern-\nulldelimiterspace} n} \leqslant 1$.
On the complex plane of the eigenvalues, the inner circle radius is ${{{(1 - c)}^{\alpha /2}}}$ and outer circle radius is unity.

\section{Experimental Results}

All the data are collected under two scenarios: (i) Only noise is present; (ii) Signal plus noise is present. We have used 70 USRP front ends and 29 high performance PCs. The experiments are divided into two main categories: (1) single USRP receiver, and (2) multiple USPR receivers. 

Every such software defined radio (SDR) platform (also called a node)  is composed of one or several USRP RF front ends and a high performance PC. The RF up-conversion and down-conversion functionalities reside in the USRP front end, while the PC is mainly responsible for baseband signal processing.
The USRP front end can be configured as either radio receiver or transmitter, which is connected with PC via Ethernet cable.

\subsection{Single USRP receiver}


For the single receiver scenario, what we observe at the USRP receiver is just a time series composed of samples of baseband waveforms.  The matrix ${\bf X} \in {\mathbb{C}^{N \times n}}$ is formed from this time series. 
Let $N=400$, $c=0.25$, $n= N/c =1600$,  as typical value. Thus the length of the time series is $N \times n = 640,000$. 

The variance and the mean of the random vector ${\bf x} \in {\mathbb{C}^{1600 \times 1}}$ are denoted as $\sigma ^{2}$ and $\mu $ respectively. The normalization is needed before we convert the time series to the corresponding random matrix $\bf X$.
Denote the $\tilde{x}_{ij}$ as the element of the raw random matrix $\tilde {\bf X} \in {\mathbb{C}^{N \times n}}$ directly composed of $N$ random realizations of the random vector ${\bf x}\in {\mathbb{C}^{1600 \times 1}}.$ Thus, the $(i,j)$-th element of the random matrix ${\bf X}$ is defined as ${x_{ij}} = \frac{{{{\tilde x}_{ij}} - \mu }}{{{\sigma}}}.$




\subsubsection{White Noise Only} 
In this scenario, only one USRP receiver is deployed and no USRP transmitter exists. We carefully select the receiving frequency to avoid any existing signal. This case is studied as the benchmark since noise always present regardless of the presence of a signal.  

\begin{figure}
 \centering
\includegraphics[height=1in]{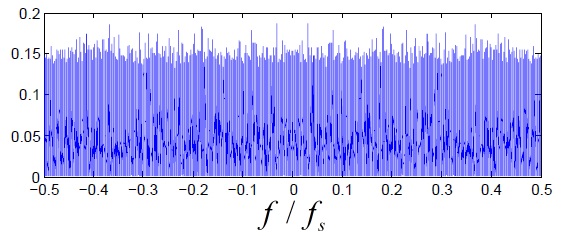}
\caption{The spectrum of the white noise captured from a USRP receiver.}
\label{whitenoise}
\end{figure}

The noise captured by the USPR receiver could be colored noise, with local oscillator leakage and DC offset. 
After removing the local oscillator leakage and DC offset, we could obtain the near white noise with spectrum shown as figure~\ref{whitenoise}.

\begin{figure}
 \centering
\includegraphics[height=1.6in]{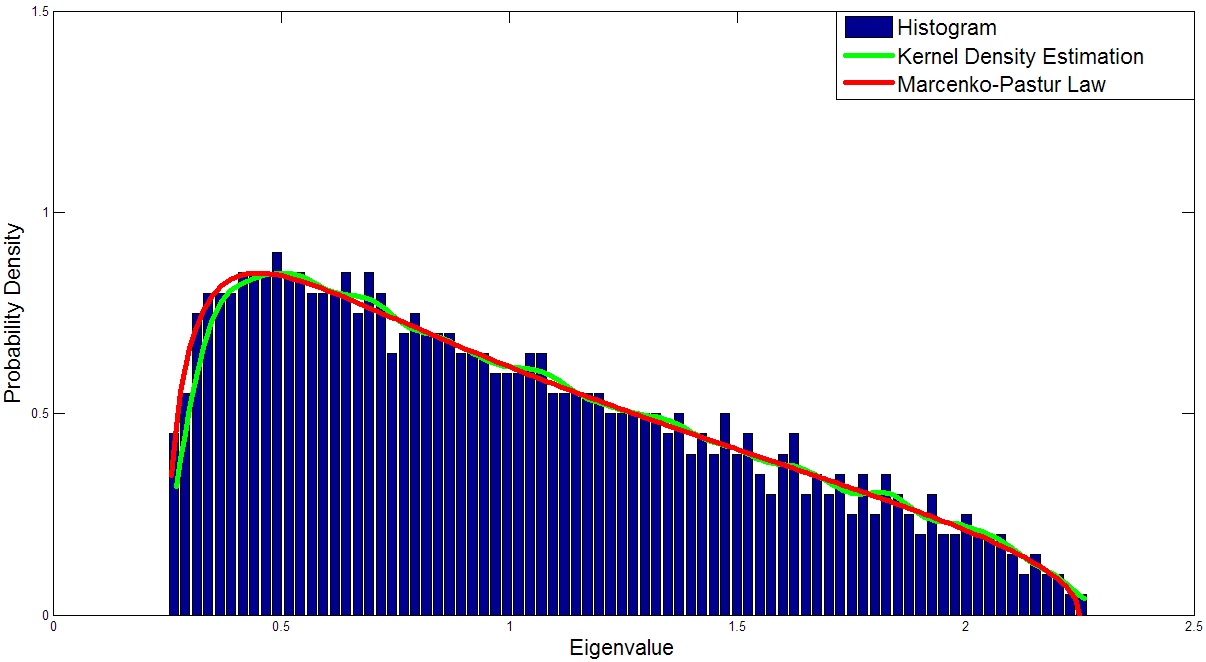}
\caption{Empirical spectrum density of white noise. The histogram (blue line)  and the kernel density estimation (green line) agree with Marchenko-Pasture Law (red line). The Marchenko-Pasture Law provides an alternative summary of relevant information in the data of Fig.~\ref{whitenoise}.}
\label{esd_whitenoise}
\end{figure}

In figure~\ref{esd_whitenoise}, we compare the histogram, kernel density estimation and the MP-law, for the ESD of the captured white noise at a single USRP receiver.
The kernel density estimation (in green) matches the histogram (in blue) very well. The most important observation is that the histogram curve and the kernel density estimation curve are very close to the MP-law. 
\subsubsection{Signal plus white noise is present}

Any deviation of from the benchmark (white noise only) will indicate the presence of ``a signal.''  We will use different types of signals to investigate such deviations.
\begin{figure}
 \centering
\includegraphics[height=1in]{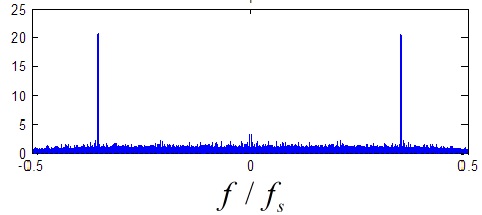}
\caption{Spectrum with commercial signal has peaks.}
\label{signal_918}
\end{figure}
\begin{figure}
 \centering
\includegraphics[height=1.6in]{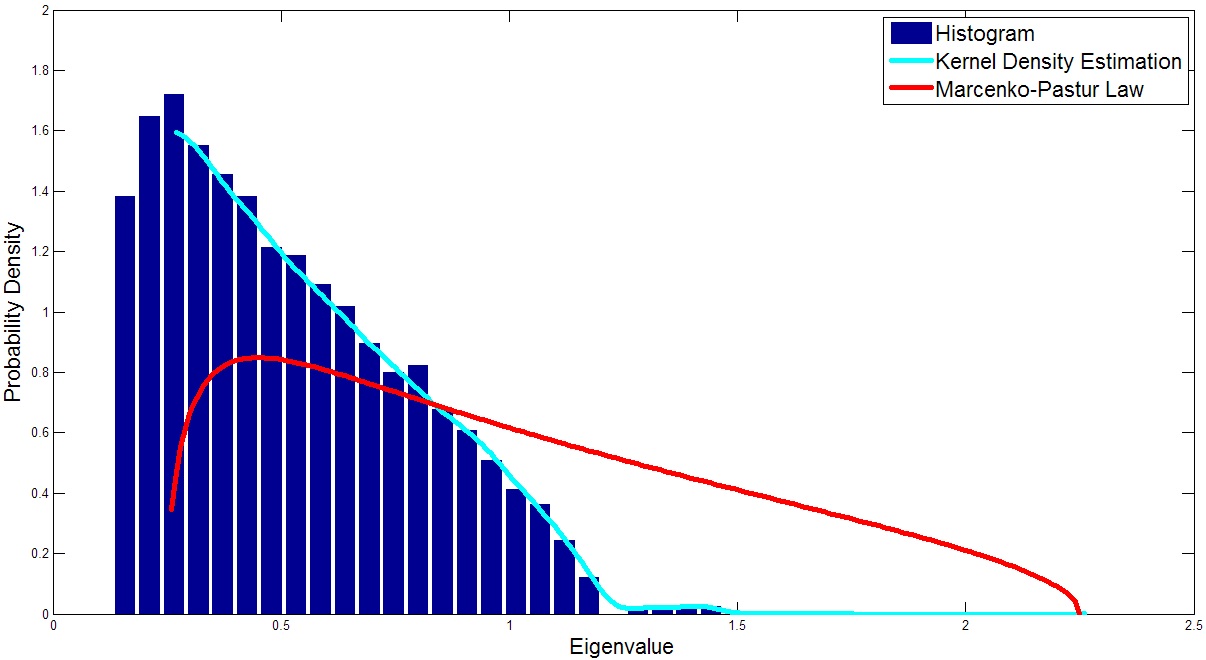}
\caption{Empirical spectrum density with commercial signal deviates from the Marchenko-Pasture law (red line). }
\label{esd_signal_918}
\end{figure}

First, we select the frequency with a commercial signal present. The base band spectrum is given in figure~\ref{signal_918}. The corresponding ESD is shown as figure~\ref{esd_signal_918}. It can be found that the ESD of the commercial signal deviates from the benchmark MP-law. The kernel density estimator still works well, since it adapts to the ``correct'' density using a kernel function.

\begin{figure}
 \centering
\includegraphics[height=1.6in]{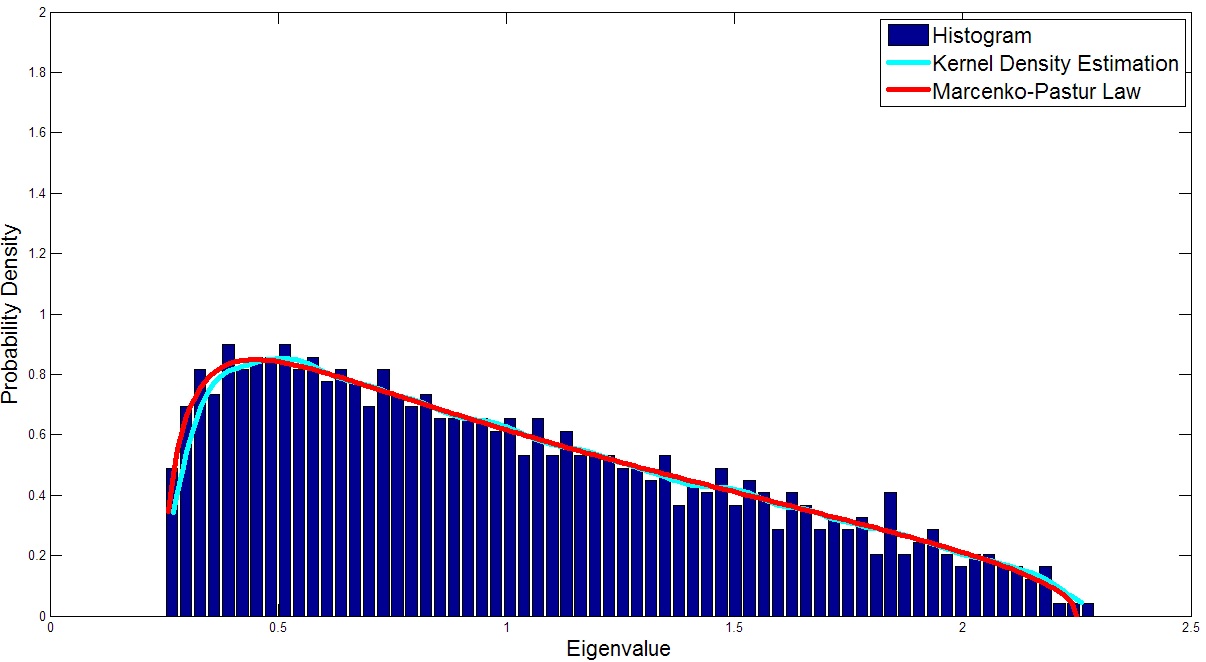}
\caption{Empirical spectrum density of the data with signal plus noise  is compliant with Marchenko-Pasture law (red line).}
\label{esd_randsig}
\end{figure}

  Second, we use another USRP to act as the radio transmitter to generate a signal with identical, independent distribution.  In this case, some random data from a video file are coded as BPSK waveforms; these waveforms are sent out without any formatting from upper layers. All the data bits follow i.i.d Bernoulli distribution. The ESD is obtained in figure~\ref{esd_randsig}. It is found that the ESD agrees with the MP-law, as the case of white noise.

	%

We summarize that the ESD always agrees MP-law if the signal or noise follows the i.i.d distribution, as predicted by the theory (Section~\ref{sect:MPlaw}). When signal samples are dependent, the ESD deviates with MP-law. In the real world, the upper layer of the communications system always add some redundant format information, like the packet header, to cause the correlation of signal samples. This observation gives an insight that the signal content can be differentiated from the noise at the waveform level by analyzing the ESD of the large random matrix.

Third, the additional question arises from approximating the ESD of the signal by a theoretical distribution, if  this ESD deviates from the benchmark MP-law. 	When fitting the data's histogram to this theoretical power law, we can adjust two parameters: (1) $\beta;$ (2) $s$ is used to scale the variance in normalization: ${x}_{ij} = s \frac{\tilde{x}_{ij} -\mu }{\sigma ^2}, s\in \left [ 0.8, 1.2 \right ].$
 	
	The strategy is to find the optimal pair of  $\left (\alpha, s  \right )$ to minimize the square error between the power law and the kernel estimation which is actually the smoothed histogram.

\begin{figure}
\centering
\includegraphics[height=1.8in]{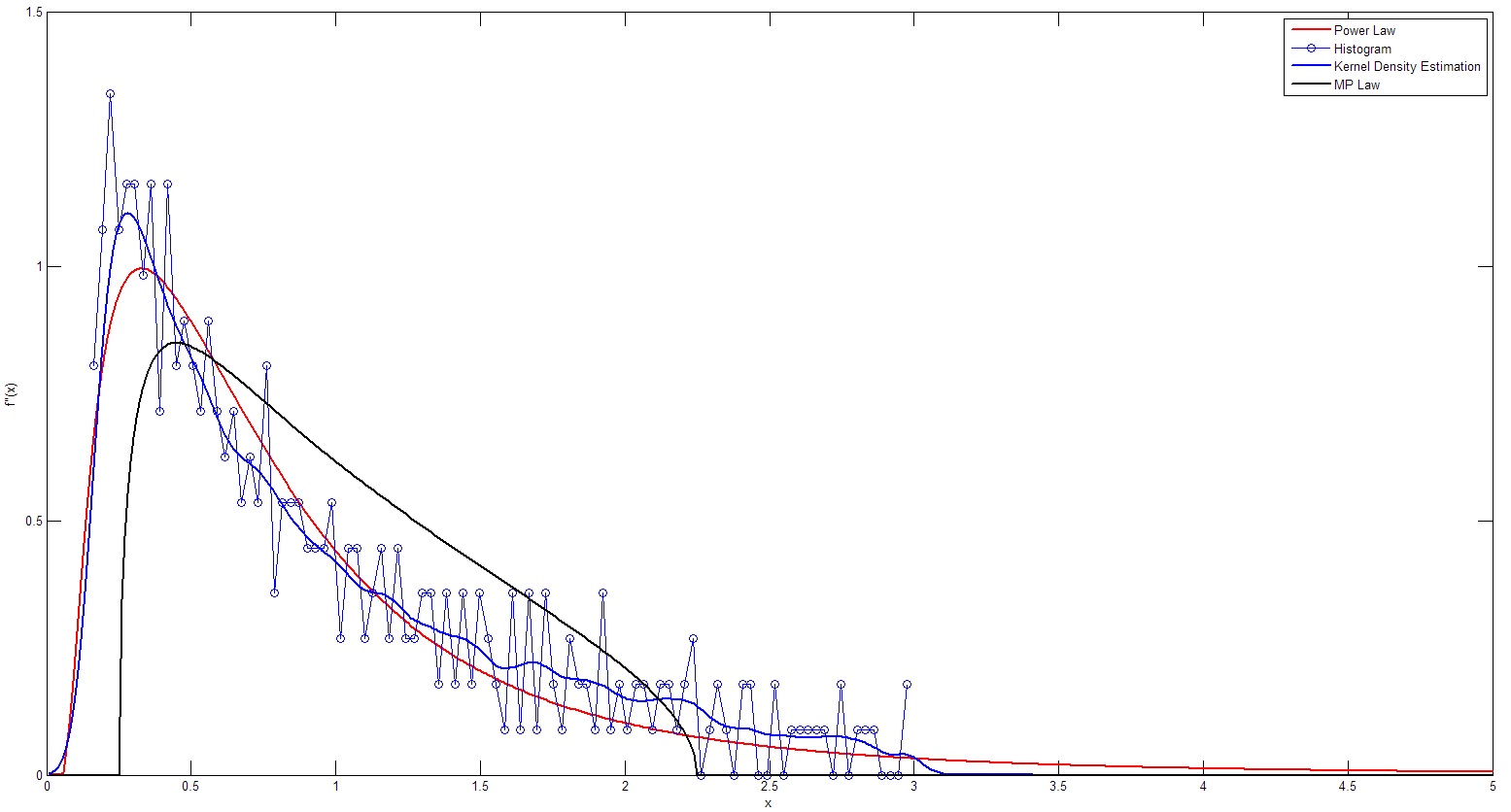}
\caption{Fitting the different power law distributions to the captured commercial signal, by adjusting some model parameters.}
\label{allinone}
\end{figure}


The USRP records a time sequence at the frequency of 869.5MHz for a WCDMA system. Figure~\ref{allinone} shows the four curves used to approximate the ESD of the recorded signal. With signal present, the MP law deviates from the histogram. By selecting proper $\alpha$ and $s$, the  power law gives a reasonably good theoretical approximation of the actual ESD.

\subsection{Multiple USRP receivers}
Seventy USRP receivers are organized as a distributed sensing network. One PC takes the role of the control node which is responsible for sending the command to all the USRP receivers that will start the sensing at the same time. The network time is synchronized by the GPS attached to every USRP. The 70 USRPs are be placed in random locations within a room. 
For every single USRP receiver, a random matrix is obtained and denoted as ${{\bf X}_i} \in {\mathbb{C}^{N \times n}}$, whose entries are normalized as mentioned above. We will investigate the ESD of the sum and the product of the $\alpha$ random matrices below, where $\alpha$ is the  number of the random matrices.

\setcounter{paragraph}{0}
\subsubsection{The Sum of the Random Matrix}
		The sum of random matrix is defined as $  \mathbf{Z} =\frac{1}{{ \sqrt \alpha   }} \sum_{i=1}^{\alpha}\mathbf{X}_{i},$ 	where $\mathbf{X}_i \in \mathbb{C}^{N\times n}$, $i = 1,\cdots,\alpha $. 
	The final random matrix is formed using the received data from $\alpha=70$ USRP receivers.
	
	
	\begin{figure}
	 \centering
	 \includegraphics[height=1.8in]{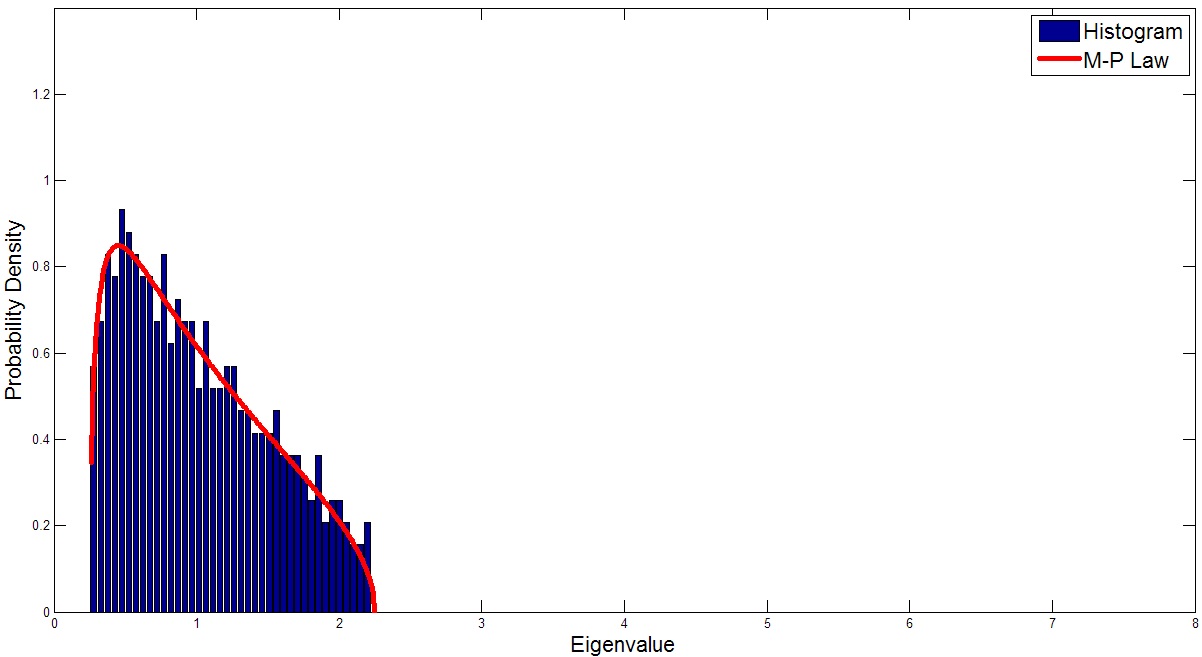}
	 \caption{The ESD of the sum of random matrices agrees with Marchenko-Pasture  law (red line). The sum of 70 normalized random matrices is formed from the pure white noise captured by 70 USRP receivers.}
		 \label{70sumnoise}
	\end{figure}
		\begin{figure}
	 \centering
	 \includegraphics[height=1.8in]{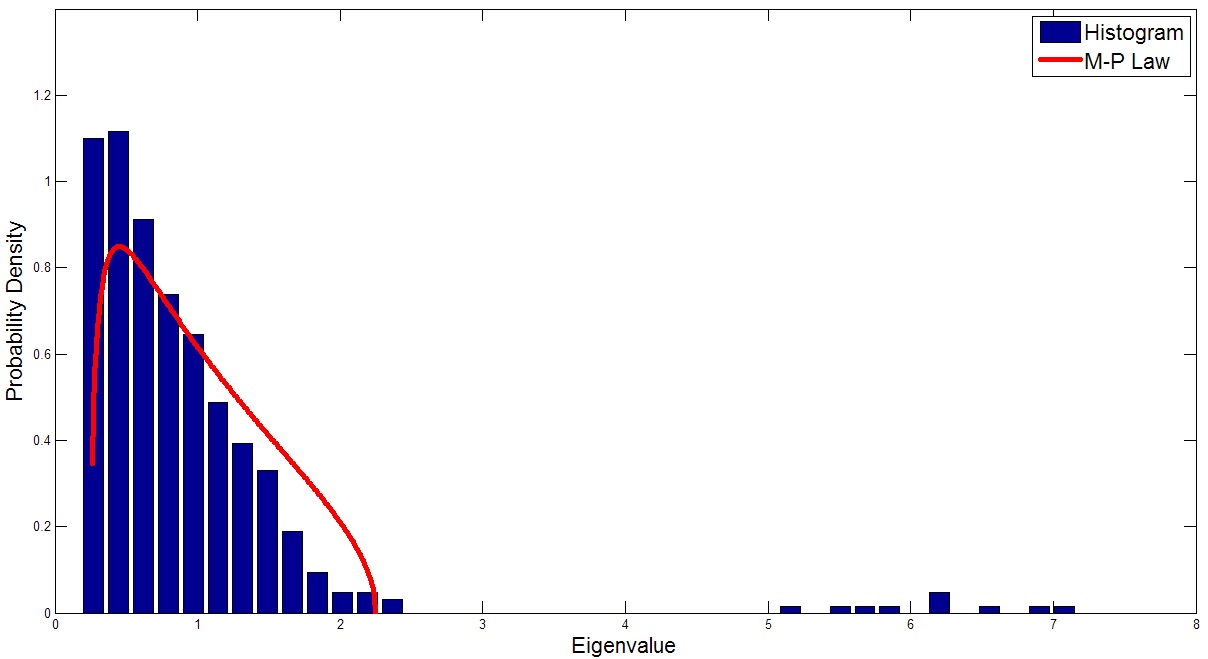}
	 \caption{The ESD of the sum of random matrices deviates from the benchmark Marchenko-Pasture  law (red line). The sum of 70 normalized random matrices is formed from the signal captured by 70 USRP receivers.}
		 \label{70sumsignal}
	\end{figure}	
	Figure~\ref{70sumnoise} shows that the ESD of the sum of the noise only random matrix agrees with the MP law. With a signal (869.5MHz commercial signal) present, Figure~\ref{70sumsignal} shows that the ESD deviates from the MP law. Some strong network ``modes'' are observed.

\subsubsection{The Product of Non-Hermitian Random Matrices}
The product of $\alpha$ non-Hermitian random matrices is defined as $	  \mathbf{Z} = \prod_{i=1}^{\alpha}\mathbf{X}_{i},$
	where $\mathbf{X}_i \in \mathbb{C}^{N\times n}$, $i = 1,\cdots,\alpha $. 
	Singular value equivalent  is performed before multiplying the original random matrices. We are actually analyzing the empirical eigenvalue distribution as~\eqref{circular_law}. 


		\begin{figure}
		\centering
		\includegraphics[height=2in]{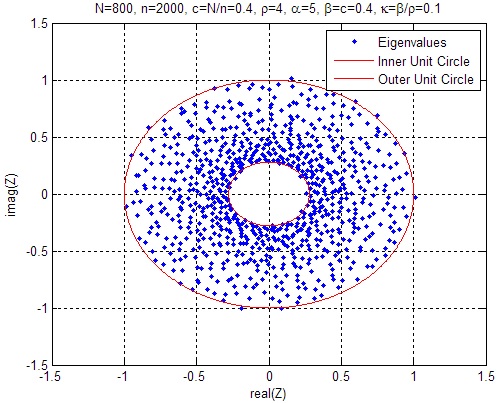}
		\caption{The ring law for the product of non-Hermitian random matrix with white noise only. The number of random matrix $\alpha$ = 5. The radii of the inner circle and the outer circle agree with~\eqref{circular_law}.}
		\label{product_circle}				
		\end{figure}
	
		\begin{figure}
		\centering
		\includegraphics[height=2in]{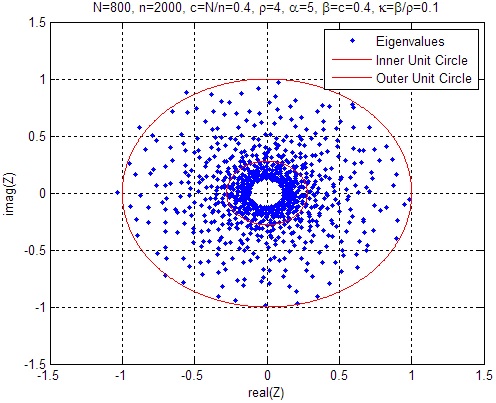}
				\caption{The ring law for the product of non-Hermitian random matrices with signal plus white noise. The number of random matrix $\alpha$ = 5. The radius of the inner circle is less than that of the white noise only scenario.}
		\label{product_circle_sig}
		\end{figure}

	By specifying $\alpha  = 5$, we performed the experiments for two scenarios:
(i) Pure Noise.  In this case, neither the commercial signal nor the USRP signal is received at all the USRP receivers.  Figure~\ref{product_circle} shows the the ring law distribution of the eigenvalues for the product of non-Hermitian random matrices when $\alpha$ is $5$. The radius of inner circle and outer circle are well matched with the result in~\eqref{circular_law}.
(ii)  Commercial Signal with Frequency. 	In this case, the signal at the frequency of 869.5MHz is used. Figure~\ref{product_circle_sig} shows the ring law distribution of the eigenvalues for the product of non-Hermitian random matrices, when signal plus white noise is present.  By comparing~\ref{product_circle_sig} with  Figure~\ref{product_circle}, we find that in the signal plus white noise case the inner radius is smaller than that of the white noise only case.


\section{Conclusion}

We have reported some initial demonstrations of the theoretical framework: the massive amount of data can be naturally represented by (large) random matrices. Although intuitive, the systematic use of this framework is relatively recent~\cite{CouilletDebbah2011BookRMT,QiuBook2012Cognitive,Qiu_WicksBook2013,QiuAntonik2014Wiley}. It appears that this work may be the first attempt to investigate the empirical science, in order to quantify the accuracy of the theoretical predictions with experimental findings. How can we exploit the new statistical information available in the massive amount of these datasets? Apparently, during this initial stage, our aim is to raise many open questions than actually answer ones. The initial results are really encouraging. One wonder if this new direction will be far-reaching in years to come toward the age of Big Data~\cite{QiuBook2012Cognitive,Qiu_WicksBook2013,QiuAntonik2014Wiley}.


    \bibliographystyle{ieeetr}   

 \begin {small}
 
 \bibliography{Bible/5GWirelessSystem,Bible/Big_Data,Bible/Compressed_Sensing,Bible/Smart_Grid,Bible/Graph_Complex_Network,Bible/Machine_Learning,Bible/Convex_Optimization,Bible/LowRankMatrixRecovery,Bible/Concentration_of_Measure,Bible/ClassicalMatrixInequalities,Bible/CompeltelyPositiveMaps,Bible/QuantumChannel,Bible/QuantumHypothesisTesting,Bible/TraceInequalities,Bible/QuantumInformation,Bible/Matrix_Inequality,Bible/RandomMatrixTheory,Bible/Fractional_Integration_bib,Bible/UWB_bib,Bible/Qiu_Group_bib,Bible/LTI_Comm_Theory_bib,Bible/Software_Defined_Radio_bib,Bible/Time_Reversal_bib,Bible/MIMO_bib,Bible/Radar_Waveform_Optim_bib,Bible/Information_Theory_bib,Bible/Compressed_Sensing_Theory,Bible/Compressed_Sensing_UWB,Bible/MISC_bib,Bible/CS_Applications,Bible/CS_Data_Stream_Algorithms,Bible/CS_Extensions,Bible/CS_Foundations_Connections,Bible/CS_Multisensor_Distributed,Bible/CS_Recovery_Algorithms,Bible/CognitiveRadio/Cognitive_Radio_bib,Bible/CognitiveRadio/CognitiveRadio2008_bib,Bible/CognitiveRadio/DySpan2007,Bible/CognitiveRadio/DySpan2005,Bible/CognitiveRadio/Gardner,Bible/CognitiveRadio/jsac200703,Bible/CognitiveRadio/jsac200801,Bible/CognitiveRadio/sensingDTV,Bible/CognitiveRadio/ucberkeley,Bible/CognitiveRadio/UWBCognitiveRadio_bib,Bible/CognitiveRadio/IEEE_JSSP_2008,Bible/CognitiveRadio/Bayesian_Network_Cognitive_Radio,Bible/CognitiveRadio/Exploiting_Historical_Spectrum,Bible/CognitiveRadio/Duke_Carin,Bible/CognitiveRadio/LiHu_upper,Bible/CognitiveRadio/AFRLref,Bible/CognitiveRadio/SVM,Bible/CognitiveRadio/NSF_ECCS_0821658,Bible/CognitiveRadio/SmartGrid}

\end {small}

%
%
%
%
%
%

\end{document}